\documentclass[a4,12pt]{article}
\usepackage{graphicx}
\usepackage{amsmath}
\usepackage{natbib}
\usepackage{geometry}[]
\newcommand{\xs}{\mathbf{x}_s}
\newcommand{\zc}{\hat{\zeta}}
\newcommand{\bk}{\mathbf{k}}
\title{Stability of a horizontal viscous fluid layer in a vertical time 
periodic electric field}
\author{Aditya Bandopadhyay  \&	  Steffen Hardt}
\date{%
Institute for Nano- and Microfluidics, \\
TU Darmstadt, \\
Alarich-Weiss-Strasse 10, 64287 Darmstadt, Germany\\%
\today}
\begin{document}
\maketitle
\begin{abstract}
The stability of a horizontal interface between two viscous fluids, one of which is conducting and the other is dielectric, acted upon 
by a vertical time-periodic electric field is considered. The two fluids are 
bounded by electrodes separated by a finite distance. 
By means of Floquet theory, the marginal stability curves are obtained, 
thereby elucidating the dependency of the critical voltage and wavenumber upon 
the fluid viscosities. 
The limit of vanishing viscosities is shown to be in 
excellent agreement with the marginal stability curves predicted by means of a 
Mathieu equation. 
The methodology to 
obtain the marginal stability curves developed here is applicable to any 
arbitrary but time periodic-signal, as demonstrated for the case of a signal with two different frequencies.  
As a special case, the marginal stability curves for an applied ac voltage 
biased by a dc voltage are depicted. 
It is shown that the mode coupling caused by the normal stress at the interface due to the electric field 
leads to appearance of harmonic modes and subharmonic modes. This is in 
contrast to the application of a voltage with a single frequency which always 
leads to a harmonic mode. Whether a harmonic or subharmonic mode is the most unstable one depends on details of the excitation signal. 
It is also shown that the electrode spacing has a distinct effect on the stability bahavior of the system. 
\end{abstract}
\tableofcontents
\section{Introduction}
The action of an electric field at the interface of two fluids having different 
properties has been a topic of extensive investigation. 
\cite{zeleny1914electrical, zeleny1917instability} considered the instability 
of such electrified liquid surfaces in the context of droplet breakup, where 
the deformation and then discharge of liquids from tubes in the presence of 
such electric-field induced stresses are considered. 
On the basis of these 
observations, \cite{taylor1964disintegration} revealed the role of the 
internal pressure and showed how the deformation and rupture of such a drop 
interface occurs. Then, based on these developments \cite{taylor1965stability} studied theoretically and experimentally the 
stability of the horizontal interface between a conducting and dielectric 
liquid upon the action of an electric field. 
By assuming a small deformation of the interface, 
they were able to obtain the conditions for neutral equilibrium at fixed 
applied dc voltages.
This anaylsis served as the basis for the study by 
\cite{yih1968stability} who considered the same physical setup as that of 
Taylor and McEwan, but where the electrodes are energized by an ac electric 
potential instead of a dc potential. 
However, in both situations, the authors assumed that the fluid is  inviscid. 
Yih showed that the stability of the interface and the height of the deformation are 
governed by the Mathieu equation, bearing an analogy with the observations 
of Faraday waves by \cite{benjamin1954stability,edwards1994patterns,kumar1994parametric}. 
The subject of coupled electric fields and fluid flow was 
then studied extensively by Melcher and coworkers \citep{melcher1963field, 
melcher1966traveling, melcher1968interfacial, jones1973dynamics}.
The fact that a rapidly varing ac field eliminates the influence of free charges 
(see c.f. 
 \cite{reynolds1965stability}) was utilized to obtain the dispersion 
relationship for the interfacial instability of initially static 
fluids \citep{devitt1965surface}.  The influence of such electric fields may also 
be employed to drive membrane fluctuations, where 
the normal stress at the interface causes a 
destabilized interface \citep{seiwert2013instability}. 

With the growing requirements set by film coating processes, there has been a 
steady 
development of techniques pertaining to pattern formation and deposition, with 
emphasis on the control 
of features on the surface (e.g. for cell manipulation, 
\cite{curtis1997topographical,ranucci2001substrate}). Such features may be 
generated by means of external dc electric fields. 
Depending on the relative competition of the 
relevant forces such as those due to surface tension and normal Maxwell 
stress, one  may obtain patterns with tunable wavelenghts. 
The methodology was utilized to generate various electrically 
tuned patterns in liquid polymer layers (the top fluid is air in such cases), 
which are then solidified by cooling below the glass transition 
temperature \citep{chou1999lithographically,chou1999lithographically2}.  
Such an instability-driven process 
was then later exploited by \cite{schaeffer2000electrically,lin2001electric} to experimentally 
obtain well-defined arrays of pillars. 
 \cite{wu2005dynamics} studied 
this phenomenon numerically for understanding the various 2D patterns such as 
hexagons, originating from the corners of the domain. 
The problems related to the action of the dc electric field on such 
interfaces have been undertaken for several configurations which are pertinent 
for various industrial applications. \cite{tilley2001dynamics} 
considered theoretically the stability of a 2D inviscid liquid film where the 
dc electric field 
is applied in the plane of the interface, and demonstrated the way in which the 
film ruptures eventually. The viscous liquid film case was considered later by 
\cite{Savettaseranee2003641, Papageorgiou2004223}. The 
interplay between the capillary pressure and the Maxwell stress was studied 
with the focus on the interfacial charge. The electric 
field is considered to be parallel to the interface, 
thus able to interact strongly when there is a 
deformation of the interface \citep{Papageorgiou200471}. The situation with a 
dc electric field normal to the interface was considered later in a separate 
study \citep{Papageorgiou2005,Papageorgiou2007832}.
Various other interfacial patterns may also be generated by 
utilizing a patterned lower electrode which affects the nature of the 
electric stresses in a spatially periodic manner
\citep{lei2003100,morariu2003hierarchical,bandyopadhyay2007electric, 
bandyopadhyay2009electric}. The process of electropatterning to obtain such 
parametrized surfaces may also be applied to multilayer systems
\citep{dickey2006novel,roberts2010electrohydrodynamic}. The combined influences of 
chemical and electric heterogeneity can also drive pattern formation in thin 
films \citep{atta2011influence, yang2014steady}. 
\cite{Tseluiko2008449, Tseluiko2008, Tseluiko2009845, 
Tseluiko2010339, Tseluiko2011169, Tseluiko2013638} considered the influence of a 
normal electric field on a thin flowing film where the surface has a topography. \cite{Wray2012430} considered the sitaution of an 
electrified film fallig down a cylinder. The work was later put into 
theoretical perspective of the electric-field modulated coating of a vertical 
fiber \citep{Wray2013427, Wray2013}.
The analysis of the 
interfacial instability in the presence of a nonuniform electric field was 
performed by \cite{yeoh2007equilibrium}.  \cite{verma2005electric} performed a 
detailed study to investigate the nature of the electric-field induced 
instabilities due to a dc field between air-liquid 
interfaces. To account for the fluid 
rheology, \cite{espin2013electrohydrodynamic}  considered the linear instability analysis for thin films 
of viscoelastic fluids for both ac and dc 
fields. Briefly, the fluid convection 
coupled to the action of electric fields (or other driving forces such as those 
due to thermal effects) may be exploited to generate micro and nanoscale patters 
(c.f. \cite{janes2013directing} for a brief review). 
The presence of a background flow has a considerable effect on the stability 
characteristics of such interfaces.  \cite{Ozen20065316} studied the 
effect of a normal electric field on a coflowing configuration of two 
immiscible fluids, where they found that the influence of the background flow 
may stabilize or destabilize the interface, depending on the physical parameters. 
This study was then followed up by \cite{Li2007347} who demarcated the 
neutral stability curves of the problem. \cite{Dubrovina2017222} studied the influence of a normal electric field on 
the pressure driven flow over a topographically modulated surface (wall 
corrugations). 
Additionally, the interested reader may also refer to the review 
articles by \cite{craster2009dynamics} and by \cite{oron1997long}.

The influence of ac fields on the interface between two fluids (a leaky dielectric drop in a leaky dielectric medium) was considered experimentally
first by \cite{torza1971electrohydrodynamic} for a forcing 
frequency of 60 Hz. 
\cite{roberts2009ac} performed a theoretical analysis of 
thin perfect dielectrics and leaky dielectrics in the presence of a vertical 
electric field. Their analysis underpins the fact that increasing the applied 
frequency leads to a stable interface. This is attributed to the fact that the 
free charge at the interface cannot react quickly enough to the rapidly changing 
electric field. This was experimentally observed by \cite{robinson2000breakdown, robinson2001electric, robinson2002nonlinear} 
who modelled the instability of the air-water interface by means of a cubic 
dissipation term in a prototypical Mathieu equation. 
Further, \cite{gambhire2010electrohydrodynamic} have 
analyzed the problem of the instability of a two-fluid interface by means of an 
ac electric field for an infinitely large domain. By focussing on the 
disturbance growth 
rate, they showed that the influence of the ac frequency has a stabilizing effect 
on the interface as compared to the dc case. By considering the situation where 
the disturbance wavelength is small, they derived the set of governing 
equations for the linear stability analysis with which they evaluated the 
Floquet multiplier. The key assumption that the two electrodes are infintiely 
spaced is not achievable in practice because to have an appreciable field 
strength, one would then require an infinitely large potential. \cite{gambhire2012role} analyzed the influence of the bottom fluid 
conductivity on the stability of the fluid interface acted upon by a normal 
electric field. The general situation with electrical double layers in both fluids was later considered by \cite{gambhire2014electrokinetic}.
Besides the action of such electric fields on planar interfaces, a lot of work 
has also been conducted towards the dynamics of jets in the presence of 
electric fields,\citep{gonzalez1989stabilization} and more importantly the 
breakup into microdrops (through a Taylor cone-like instability).
\cite{li2009transient} studied the influence of the electric Weber 
and Euler numbers towards estimating the transient growth rate for two- 
and three-dimensional disturbances.  \cite{mandal2015electro} 
considered the instability at the two-phase aqueous interface in the presence of 
charge modulated surfaces and a longitudinal applied dc electric field. 
\cite{navarkar2016long} performed a longwave linear stability analysis 
for a two-phase aqueous solution system with focus on the nature of 
stability with regard to the interfacial charge. \cite{Conroy2011} 
studied theoretically how ac fields lead to the breakup of a viscous thead, in 
particular how the field strength may be tuned to control the number of 
satellite drops.  

In the present analysis, we consider the interface between two viscous fluids 
acted upon by a time periodic-electric field. No assumption about the disturbance wavelength is made, i.e. we do not assume long wavelength 
disturbances. 
Furthermore, we consider the realistic situation that the electrodes are 
separated by a finite distance. 
An infinitely extended fluid interface is assumed, 
i.e. all wavelengths are permitted. The problem under these 
realistic conditions has not been addressed in literature, to the best of the 
authors' knowledge. 
In our approach, the deformed interface acts as a domain perturbation to 
the system, and we obtain the set of 
equations which govern the normal velocity component and surface deformation. 
By means of Floquet theory we obtain the marginal stability curves of the 
problem for a generalized form of the applied ac voltage, i.e. there can be 
multiple frequencies present. Such a generalized form is central to 
representing any arbitrary applied signal. 
In the limit of vanishingly small viscosities we show 
that our results 
reduce to those reported by \cite{yih1968stability} through the Mathieu 
equation. 
The structure of the marginal stability curves reveals the fact 
that, in general, the ac frequencies serve to dampen the instabilities. 
In context with multiple applied frequencies it is shown that 
because the stress due to the electric field is proportional to $V(t)^2$, the interaction 
among different modes leads to the occurence of zones of harmonic and 
subharmonic surface deformation modes corresponding to different wavenumbers. Finally, the effect 
of the distance between the two electrodes on the critical voltage 
and wavenumber is considered. It is hoped that that procedure here would help in understanding the pattern formation for multiple applied frequencies and finetuning observed wavelengths for fabricating intricate structures.

\section{Mathematical formulation}
\subsection{Setup}

\begin{figure}
\centering
\includegraphics[scale=1]{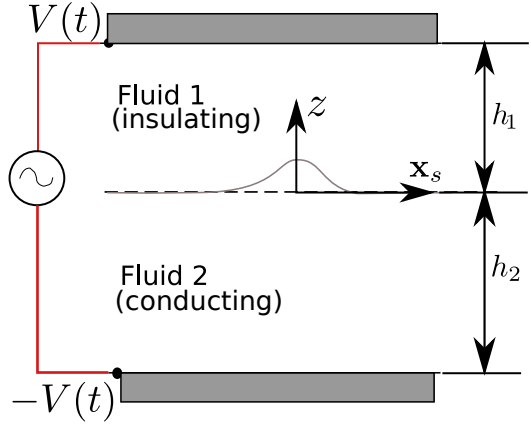}
\caption{The system comprises of a lighter fluid, fluid $1$, and a heavier 
fluid, fluid $2$. The top electrode, $z = h_1$, and the bottom electrode, $z = 
-h_2$, are energized by means of an ac voltage, $V(h_1, -h_2) = (V(t), -V(t))$. 
The lower fluid conducting while the top is insulating (nonconducting). 
The interface deforms due to the influence of the electric field manifested in terms of the Maxwell stress. The 
coordinate system is fixed to the undeformed interface.}
\label{Fig:schematic}
\end{figure}

Consider the situation where a lighter non-conducting fluid is on top of a 
heavier conducting fluid. The system of the two fluids is bounded by two electrodes. Such a system has been considered by 
\cite{taylor1965stability} for the case of a dc electric field acting on 
a pair of non-conducting/conducting inviscid fluids and later by 
\cite{yih1968stability} for the same fluids under an ac electric field (see figure \ref{Fig:schematic}).
As an example, the bottom fluid could be aqueous salt solution and the top fluid 
vegetable oil (castor oil, corn oil, canola oil etc.) or a synthetic oil 
(silicone oil, mineral oil, etc.). 
The subscripts 1 and 2 for the properties below denote the top and bottom fluid, 
respectively. 
The properties of the $i$th fluid are: density, $\rho_i$, viscosity, $\eta_i$, 
and permittivity, $\epsilon_i$. 
The coordinate system is attached to the undeformed surface, where $z$ 
represents the coordinate normal to the undeformed interface and $\xs \left(=  
x, y \right)$ represents the surface coordinates. 
We consider an infinitesimal deformation of the interface given by 
\begin{eqnarray}
z = \zeta(\xs,t) = \zc(t) \sin \bk\cdot \xs,
\end{eqnarray}
where $\zc(t)$ represents the time-dependent amplitude and $\bk = (k_x, k_y)$ 
represents the surface wavevector. The normal vector, $\mathbf{n}$, and 
curvature, $\kappa$, corresponding to this deformation are given by 
\begin{eqnarray}
\mathbf{n} = \frac{\nabla (z-\zeta(\xs,t))}{|\nabla (z-\zeta(\xs,t))|} \approx
\mathbf{e}_z - \nabla_s \zeta(\xs,t), \qquad \kappa = \nabla\cdot \mathbf{n} = 
-\nabla_s^2 \zeta(\xs,t),
\end{eqnarray}
where $\nabla_s$ represents the surface derivative operator. 
\subsection{Description of the electric field}
The governing equation for the electric potential in the top fluid is given by 
\begin{eqnarray}
\nabla^2\psi_1(\xs,z,t) = 0.
\label{eq:potential_equation}
\end{eqnarray}
The consideration of a conducting lower fluid implies that the potential in the 
lower fluid is equal to that applied at the lower electrode. 
Following the convention by \cite{yih1968stability}, the boundary conditions for the 
applied electric field are 
\begin{eqnarray}
\psi_1(\xs,h_1,t) = -V(t), \qquad
\psi_1(\xs,\zeta(\xs,t),t) = V(t),
\end{eqnarray}
where $\psi_i$ is the electric potential in the $i$th fluid, $h_1$ and 
$-h_2$ represent the locations of the top and bottom electrode, and $V(t)$ 
is the applied ac voltage. For the special case of an applied voltage 
at a single frequency, we obtain the case corresponding to that of Yih, i.e. 
$V(t)= -V \cos(\omega t)$. However, we retain the general form of the applied 
voltage for convenience. The potential in the lower fluid does not need to be solved for 
since it is conducting \citep{taylor1965stability, yih1968stability}.

We can employ a regular perturbation approch in terms of $\zc$ to represent the potential 
as 
\begin{eqnarray}
\psi_1(\xs,z,t) = \psi_1^{(0)} (\xs,z,t) + \zc \psi_1^{(1)}(\xs,z) + O(\zc^2).
\label{eq:pot_perturbation}
\end{eqnarray}
Substituting equation \eqref{eq:pot_perturbation} in equation 
\eqref{eq:potential_equation} we obtain 
\begin{eqnarray}
\nabla^2 \psi_1^{(0)} = 0, \qquad \nabla^2 \psi_1^{(1)} = 0,
\label{eq:potential_equations}
\end{eqnarray}
where we have implicitly assumed that the electric potential is quasistationary 
which is true for processes occurring slower than the electric relaxation 
timescale, $\epsilon_2/\sigma_2$ \citep {melcher1969electrohydrodynamics, 
saville1997electrohydrodynamics}.
The boundary conditions on the deformed surface may be written in terms of conditions for the flat 
interface by means of the Taylor series expansion about $z= 0$, as follows:
\begin{equation}
\begin{split}
\psi_1|_{z=\zeta} &= \psi_1^{(0)}|_{z=0} + \zc \psi_1^{(1)}|_{z=0} + 
\zeta \left( \frac{\partial \psi_1^{(0)}}{\partial z} \right)\vert_{z=0} + 
O(\zeta^2)
\\
&= \psi_1^{(0)}|_{z=0} + \zc \sin \bk\cdot\xs \left( \frac{\partial 
\psi_1^{(0)}}{\partial z} + \hat{\psi}_1^{(1)}\right)\vert_{z=0} + ...,
\end{split}
\label{eq:potential_bcs}
\end{equation}
where we have made use of the decomposition $\psi_1^{(1)}(\xs,z,t) = 
\hat{\psi}_1^{(1)}(z,t)\sin \bk\cdot \xs$.

Proceeding further, we note that the leading order solution for the electric potential, $\psi_1^{(0)}$, obeying the pertinent boundary condition on the flat 
interface (the O(1) boundary condition from equation \eqref{eq:potential_bcs}), 
is only a function of the $z$ coordinate. Therefore the solution for the electric potential at leading 
order is given by 
\begin{eqnarray}
\psi_1^{(0)}(z,t) = V(t) - 2 V(t)\frac{z} {h_1}.
\label{eq:zeta_0}
\end{eqnarray}

The O($\zc$) electric potential, $\psi_1^{(1)}$ may be obtained 
by utilizing the aforementioned decomposition of the potential to obtain the following equation for $\hat{\psi}_1^{(1)}$ and the following boundary conditions:
\begin{gather}
\frac{d^2 \hat{\psi}_1^{(1)}}{dz^2} - k^2 \hat{\psi}_1^{(1)} = 0, \\
\hat{\psi}_1^{(1)}(z=h_1) = 0, \qquad \hat{\psi}_1^{(1)}(z=0) = 
-\frac{d\psi_1^{(0)}}{dz} = \frac{2V(t)}{h_1},
\label{eq:ozeta_pot}
\end{gather}
where $k^2 = |\bk|^2$ represents the magnitude of the wavevector. 
Equation \eqref{eq:ozeta_pot} is solved to obtain 
\begin{eqnarray}
\hat\psi_1^{(1)} = \frac{2V(t)}{h_1} \frac{\sinh(k(h_1-z)}{\sinh(kh_1)}.
\label{eq:zeta_1}
\end{eqnarray}
Consequently, one may write the total solution for the electric potential up to order $\zc$ by combining 
equations \eqref{eq:zeta_0} and \eqref{eq:zeta_1} together with equation 
\eqref{eq:pot_perturbation} to obtain:
\begin{eqnarray}
\psi_1 = V(t)\left( 1- \frac{2z}{h_1}\right) + \zc(t) \sin \bk\cdot \xs 
\frac{2V(t)}{h_1}\frac{\sinh(k(h_1-z)}{\sinh(kh_1)}.
\end{eqnarray}
This result may also be derived equivalently by assuming a surface disturbance of the form $z = \zc(t)f(x,y) $ with $f(x,y)$ satisfying the 
Laplace equation (see  \cite{taylor1965stability} and 
\cite{yih1968stability}). A function of the form $f(x,y) = \sin \bk\cdot \xs = \sin (k_xx+k_yy)$ fulfils these requirements. 

The normal electric stress is obtained by noting that 
\begin{eqnarray}
\mathbf{n}\cdot\tau^{E}\cdot\mathbf{n} = 
\mathbf{e}_z\cdot\tau_{zz}^{E}\mathbf{e}_z -2\zc \cos\bk\cdot\xs 
\bk\cdot\tau_{zz}^{E}\mathbf{e}_z + O(\zc^2)
\end{eqnarray}
where $\tau^{E}$ represents the Maxwell stress tensor. 
The above consideration yields the leading order stress at the interface, $z = 
\zeta(\xs,t)$ as follows. The electric field in the $z$ direction is obtained as:
\begin{eqnarray}
E_z(\zeta) = -\frac{\partial \psi_1}{\partial z} = \frac{2V(t)}{h_1} + \zc \sin 
\bk\cdot\xs \frac{2V(t)k}{h_1}\frac{\cosh k(\zeta-h_1)}{\sinh kh_1}
\end{eqnarray} 
which, when written as a Taylor series about $z=0$ as $E_z(\zeta) = E_z|_{0} + 
\frac{\partial E_z}{\partial z}\zeta(\xs,t) $, only contributes through the 
first term because the second term is $O(\zc^2)$. The electric field in the 
lower conducting fluid is zero. This allows us to obtain 
the jump in the normal electric stress as 
\cite{melcher1969electrohydrodynamics} 
\begin{eqnarray}
\frac{1}{2} \epsilon_1 E_z^2 =2\epsilon_1 \left( \frac{V}{h_1}\right)^2 + \zc 
\sin \bk\cdot\xs\left(4\frac{\epsilon_1 V^2k}{h_1^2}\coth kh_1 \right).
\label{eq:normal_electric_stress}
\end{eqnarray}

\subsection{Hydrodynamics}
We now turn our attention to the hydrodynamics of the system. The two fluids 
satisfy the Navier Stokes equation for a Newtonian fluid:
\begin{eqnarray}
\rho_j \left( \frac{\partial}{\partial t} + \mathbf{u}_j \cdot \nabla\right) 
\mathbf{u}_j  = -\nabla P_j + \eta_i \nabla^2 \mathbf{u}_j -\rho_j g \mathbf{e}_z
\label{eq:hydro},
\end{eqnarray}
where $j=1,2$ represents the two fluids. We then consider the representation of the 
pressure and velocity given by the base state and perturbation terms as
\begin{equation}
\begin{split}
\mathbf{u}_j(\xs,z,t) &=  \zc(t) \mathbf{\hat{u}}_j(\xs,z) + O(\zc^2),\\
P_j(\xs,z,t) &= P_j^{(0)}(z) + p_j^{(1)}(\xs,z,t) + O(\zc^2),
\end{split}
\label{eq:v_pert}
\end{equation}
which implies that the leading-order velocity field is zero. 
Equation \eqref{eq:v_pert} may be substituted in equation \eqref{eq:hydro} and 
split into various orders $O(\zc)$. 
The base state hydrostatic solution is given by 
\begin{eqnarray}
P_1^{(0)} = -\rho_1 g z + C_1, \quad P_2^{(0)} = -\rho_2 g z + C_2,
\end{eqnarray}
At $O(\zc)$ the governing equations for the fluids are obtained as
\begin{equation}
\rho_j \frac{\partial \mathbf{u}_j}{\partial t}  = -\nabla p_j^{(1)} + \eta_j 
\nabla^2\mathbf{u}_j, \qquad j = 1,2.
\end{equation}
We can then apply the operator $\mathbf{e}_z\cdot\nabla\times\nabla$ to the momentum equation to obtain 
\begin{eqnarray}
\left( \frac{\partial}{\partial t} - \nu_i \nabla^2\right)\nabla^2 w_j = 0, 
\qquad j=1,2
\label{eq:eqn_w}
\end{eqnarray}
where $w_j$ represents the $z$-component of the velocity field. Equations \eqref{eq:eqn_w} are subjected to the following boundary conditions: 
\begin{eqnarray}
\text{at}\, z= h_1,\; \mathbf{u}_{s,1} = 0, \implies  w_1 = 0, \frac{\partial 
w_1}{\partial z} = 0, \\
\text{at}\, z= -h_2, \; \mathbf{u}_{s,2} = 0, \implies w_2 = 0,  \frac{\partial 
w_2}{\partial z} = 0, 
\end{eqnarray}
while at the interface, we have 
\begin{eqnarray}
\mathbf{u}_1 = \mathbf{u}_2 \implies w_1 = w_2, \; \frac{\partial w_1}{\partial 
z} = \frac{\partial w_2}{\partial z}, 
\end{eqnarray}
which is to be evaluated at $z = \zeta(\xs,t)$ and may be expanded about $z= 0 $ by 
means of a Taylor series. It must be reiterated that the velocity itself is 
$O(\zc)$ (by virtue of the base velocity being zero). 
The kinematic boundary condition may be written as 
\begin{equation}
w_1(\xs,\zeta,t) = w_2(\xs,\zeta,t) = \frac{\partial \zeta(\xs,t)}{\partial t} 
+ \mathbf{u}_s \cdot \zeta(\xs,t).
\end{equation}
The tangential stress boundary condition yields: 
\begin{eqnarray}
\eta_1 \left( \nabla_s^2 w_1 + \frac{\partial }{\partial z} \nabla_s \cdot 
\mathbf{u}_{s,1}\right) = \eta_2 \left( \nabla_s^2 w_2 + \frac{\partial 
}{\partial z} \nabla_s \cdot \mathbf{u}_{s,2}\right), 
\end{eqnarray}
where $\mathbf{u}_{s,i} = (u_i,v_i)$ represents the surface velocity vector of 
the $i$th fluid. We note that the contribution of the Maxwell stress tensor to 
the tangential stress is zero, as a conducting interface cannot sustain any free charge, i.e. $\mathbf{t}\cdot\tau^{E}\cdot\mathbf{n} = 0$ at $z = \zeta$, where $\mathbf{t}$ represents the tangential direction \citep{melcher1963field}. This is a consequence of the fact that at the interface of the conducting fluid, $\mathbf{n}\times \mathbf{E} = \mathbf{0}$, i.e. the electric field is normal to the interface. 
The above hydrodynamic stress balance arises from taking a surface divergence of the viscous stress
$\tau_{\xs z}$ for both the fluids. 
The final dynamic boundary condition at the interface stems from the normal 
stress balance at $O(\zc)$. The normal stress boundary condition at $z=\zeta$ 
yields
\begin{equation}
\begin{split}
\left(-P_1+2\eta_1\frac{\partial w_1}{\partial z}\right) - \left( -P_2 + 
2\eta_2\frac{\partial w_2}{\partial z}\right) + 2\epsilon_1\left( 
\frac{V}{h_1}\right)^2 + \zeta \left( 4\frac{\epsilon_1 V^2 k}{h_1^2}\coth kh_1 
\right) = -\sigma \nabla_s^2 \zeta,
\end{split}
\end{equation}
\begin{equation}
\left( 
-p_1^{(1)} + \rho_1 g\zeta+ 2\eta_1 \frac{\partial w_1}{\partial z}
\right) - 
\left(
-p_2^{(1)} + \rho_2 g\zeta + 2\eta_2 \frac{\partial w_2}{\partial z} 
\right) 
+ \zeta \frac{4\epsilon_1 V^2k}{h_1^2}\coth kh_1 
= -\sigma \nabla_s^2 \zeta, 
\label{eq:normal_stress_zeta}
\end{equation}
where we have used the fact that $-C_1 + C_2 + 2\epsilon_1\left( 
\frac{V}{h_1}\right)^2 = 0$ which represents the leading-order balance of the pressure and normal Maxwell stress. This consideration leads to equation \eqref{eq:normal_electric_stress}, wherein we note that each term is 
$O(\zc)$.
Simplifying the above, we obtain at $z = 0$
\begin{equation}
p_1^{(1)} - p_2^{(1)} = 2(\eta_1-\eta_2)\frac{\partial w}{\partial z} + \sigma 
\nabla_H^2\zeta + (\rho_1 - \rho_2)g \zeta + \frac{4\epsilon_1 
V^2k}{h_1^2}\coth kh_1 \zeta.
\label{eq:delta_p}
\end{equation}
The left-hand side of the above equation may be rewritten by making use of the 
momentum equation (applying the surface-divergence operator to the $\xs$-momentum 
equations):
\begin{equation}
\nabla_s^2p_i = \left( \rho_i \frac{\partial }{\partial t} - \eta_i \nabla^2 
\right) \frac{\partial w_i}{\partial z},
\label{eq:nabla2_p}
\end{equation}
where we have used the continuity equation, $\nabla_s\cdot \mathbf{u}_{s,i} + 
\partial w_i/dz=0$.
Thus, substituting equation \eqref{eq:delta_p} in equation \eqref{eq:nabla2_p}, we obtain 
\begin{equation}
\begin{split}
\nabla_s^2 \left( \Delta\rho g \zeta +  2\Delta\eta\frac{\partial w}{\partial 
z} + \sigma \nabla_s^2 \zeta + \frac{4\epsilon_1 V^2 k }{h_1}\coth kh_1 \zeta 
\right) \\= \rho_1 \frac{\partial^2 w_1}{\partial t\partial z}  -  \rho_2 
\frac{\partial^2 w_2}{\partial t\partial z}- \eta_1\nabla_s^2 \frac{\partial 
w_1}{\partial z} + \eta_2\nabla_s^2\frac{\partial w_2}{\partial z}.
\end{split}
\label{eq:normalstress}
\end{equation}
The normal stress boundary condition yields the necessary closure required to 
perform the stability analysis. 

\subsection{Floquet theory}

The decomposition of the $z$-velocity and deformation appearing in the normal 
stress boundary condition is represented by
\begin{equation}
(w_i,\zeta) = (\hat{w}_i(z,t), \zc) \sin \bk\cdot\xs. 
\end{equation}
We shall now apply Floquet theory to determine the stability of the system. 
Towards this, we assume that the solutions of the out-of-plane components, $\hat{w}_i, \hat{\zeta}$, are of the from 
\begin{gather}
\hat{w}_j(z,t) = \exp(st+i\alpha\omega t)W_j(z,t \text{mod} 2\pi/\omega) =  
\exp(st+i\alpha\omega t) \sum\limits_{n=-\infty}^{n=\infty} W_{j,n} 
\exp{in\omega t},\\
\zc = \exp(st+i\alpha\omega t) \sum\limits_{n=-\infty}^{n=\infty} Z_{n} 
\exp{in\omega t},
\end{gather}
where $\exp(st + i\omega\alpha t)$ represents the nonunique Floquet 
(characteristic) multiplier, and the summation part represents a general time-periodic term which is decomposed into an infinite series of Fourier modes. The Floquet multiplier is nonunique because of the fact that for a given $s$, $s+2\pi\omega j$ ($j$ is an integer) is also a solution. When the forcing frequency is 
$\omega$, the constraint for $\alpha$, $0\leq\alpha \leq 1/2$, renders it unique 
with $\alpha = 0$ and $1/2$ representing the harmonic and the subharmonic 
response of the system. respectively (\cite{kumar1994parametric}). 
Substituting these forms in the governing equations, we obtain 
\begin{eqnarray}
\left( s+i(\alpha + n)\omega - \nu_1\left( \frac{d^2 }{dz^2}- k^2\right) 
\right) \left( \frac{d^2 }{dz^2}-k^2 \right) W_{1,n} = 0,\\
\left( s+i(\alpha + n)\omega - \nu_2\left( \frac{d^2 }{dz^2}- k^2\right) 
\right) \left( \frac{d^2 }{dz^2}-k^2 \right) W_{2,n} = 0.
\label{eq:govern_velo_modes}
\end{eqnarray}
To simplify the following expressions, we introduce the abbreviations 
\begin{eqnarray}
q_n^2 = k^2 + \frac{s + i (\alpha + n)\omega}{\nu_1} \\
r_n^2 = k^2 + \frac{s + i (\alpha + n)\omega}{\nu_2}.
\end{eqnarray}
Utilizing this, we obtain the solution for the vertical velocity modes as
\begin{eqnarray}
W_{1,n} = P_{1,n} \exp (kz) + Q_{1,n}\exp (-kz) + R_{1,n}\exp (q_n z) + S_{1,n} 
\exp (-q_n z) \\
W_{2,n} = P_{2,n} \exp (kz) + Q_{2,n}\exp (-kz) + R_{2,n}\exp (r_n z) + S_{2,n} 
\exp (-r_n z).
\label{eq:W_comp_sol}
\end{eqnarray}
The constants $(P_{i,n}, Q_{i,n}, R_{i,n}, S_{i,n})$ are obtained by utilizing 
the following boundary conditions:
\begin{eqnarray}\label{eq:bcs_1}
W_{1,n}(h_1) = 0, \;\; \frac{\partial W_{1,n}}{\partial z}(h_1) = 0, \\
W_{2,n}(-h_2) = 0, \;\; \frac{\partial W_{2,n}}{\partial z}(-h_2) = 0, \\
W_{1,n}(0) = W_{2,n}(0),\;\; \frac{\partial W_{1,n}}{\partial z}(0) = 
\frac{\partial W_{1,n}}{\partial z}(0), \\
\eta_1\left( k^2 + \frac{\partial^2 }{\partial z^2} \right)W_{1,n} = 
\eta_2\left( k^2 + \frac{\partial^2 }{\partial z^2} \right)W_{2,n}
\end{eqnarray}
\begin{eqnarray}
 \left( q_n^2 - k^2 \right) \nu_1 Z_n = W_{1,n}.
 \label{eq:bcs_2}
\end{eqnarray}
We can simplify matters here by the assumption of a large separation between the electrodes,  $(h_1,-h_2)\rightarrow (\infty, -\infty)$. This allows us to write
the solution for the velocity field (equation \eqref{eq:W_comp_sol}) in a simplified manner as 
\begin{eqnarray}
W_{1,n} = Q_1\exp{(-kz)} + S_1\exp(-q_nz)\\
W_{2,n} = P_2\exp(kz) + R_2\exp(r_nz),
\label{eq:solution_W}
\end{eqnarray}
which takes care of the boundary conditions at $z=h_1,h_2$  by construction. 
The general case of finite electrode spacings is described in section \ref{sec:finite_h}.
Thus, utilizing the conditions at the interface, we have 
\begin{gather}
Q_{1,n} + S_{1,n} = P_{2,n} + R_{2,n}\\
-kQ_{1,n} -q_n S_{1,n} = kP_{2,n} +r_n R_{2,n}\\
\eta_1\left( 2Q_{1,n}k^2+S_{1,n}q_n^2+k^2S_{1,n}\right) = 
\eta_2\left( 2P_{2,n}k^2 + R_{2,n}r_n^2 + k^2R_{2,n}\right)\\
(q_n^2-k^2)\nu_1Z_n = Q_{1,n}+ S_{1,n},
\end{gather}
which is solved to obtain 
\begin{gather}
S_{1,n} = -\frac{2\nu_1k(q_n+k)(k\eta_1+r_n\eta_2)}{(k+1_n)\eta_1+ 
(k+r_n)\eta_2}Z_n\\
Q_{1,n} = 
\frac{\nu_1(k+q_n)((\eta_1-\eta_2)k^2+\eta_2(q_n+r_n)k+\eta_1q^2+\eta_2r_nq_n)}{
(\eta_1+\eta_2) k+\eta_2r_n+\eta_1 q_n}Z_n\\
P_{2,n} = 
\frac{(k+q_n)(k-q_n)((\eta_1-\eta_2)k^2-\eta_1(q_nk+r_nk+q_nr_n)-\eta_2k^2q_n^2)
}{((\eta_1+\eta_2)k+\eta_2r+q\eta_1)(r_n-k)}Z_n\\
R_{2,n} = 
\frac{2\nu_1k(q_n^2\eta_2k+q_n^2\eta_1-\eta_2k^3-k^2\eta_1q)}{
(k-r_n)(\eta_1(k+q_n)+\eta_2(k+r_n))}Z_n.
\end{gather}

\subsection{Discrete representation}
Having obtained the solution for the various modes of $w_1$ and $w_2$ in 
terms of $Z_n$, we can proceed to utilize the normal stress balance to obtain a 
discrete representation of the problem in the form of an eigenvalue problem to determine the marginal stability curves. 
Referring to equation \eqref{eq:normalstress}, we see that all the terms except 
the normal electric stress are linear and have no mode coupling. The mode 
coupling is responsible for alterations in the marginal stability due to the 
influence of the electric field. 
By making use of the representation of the velocity and displacement through the Floquet form multiplied by the Fourier modes, we can 
represent the normal stress balance as a discrete problem as shown below. Focusing first on the normal electric stress without the prefactor,  we see that
\begin{equation}
V(t)^2\zeta = V(t)^2\sum\limits_{n=-\infty}^{n=\infty}\exp((s+i(\alpha+n)\omega 
t)) Z_n
\end{equation}
which can be further simplified by knowing the form of $V(t)$. For 
example, for the situation of a single mode ac electric field, we have 
\begin{eqnarray}
V(t)^2\zeta &= V_0^2\sum\limits_{n=-\infty}^{n=\infty}
\left(\frac{\exp(i\omega t) + \exp(-i\omega t)}{2}\right)^2\exp((s+i(\alpha+n) 
\omega t)) Z_n \\
&= V_0^2\sum\limits_{n=-\infty}^{n=\infty}
\left(\frac{\exp(2i\omega t) + \exp(-2i\omega 
t)+2}{2}\right)\exp((s+i(\alpha+n) \omega t)) Z_n 
\end{eqnarray}
\begin{equation}
=\frac{ V_0^2}{4} \sum\limits_{n=-\infty}^{n=\infty} \left( 
\exp(s+i(\alpha+n+2) \omega t) \\
+ \exp(s+i(\alpha+n-2) \omega t) + 
2\exp(s+i(\alpha+n) \omega t)\right) Z_n,
\end{equation}
where the first and second term can be recast in terms of $Z_{n-2}$ and 
$Z_{n+2}$ to obtain a prefector of $\exp((s+i(\alpha+n) \omega t))$.
Upon combining this term with the other terms arising in the normal stress 
balance, we obtain

\begin{equation}
\begin{aligned}
\rho_1 \nu_1\left( q_n^2 - k^2 \right) \frac{\partial W_{1,n}}{\partial z} - 
\rho_2 \nu_1\left( r_n^2 - k^2 \right) \frac{\partial W_{2,n}}{\partial z} \\
-
\eta_1 \left( \frac{\partial ^2}{\partial z^2} - k^2 \right) \frac{\partial 
W_{1,n}}{\partial z} +
\eta_2 \left( \frac{\partial ^2}{\partial z^2} - k^2 \right) \frac{\partial 
W_{2,n}}{\partial z} + 
 2(\eta_1-\eta_2)k^2\frac{\partial W_{1,n}}{\partial z}\\
= -g (\rho_1 - \rho_2)Z_n + \sigma k^4 Z_n - 
\epsilon_1\frac{4V_0^2}{h_1}k^3\coth 
kh_1 (Z_{n+2}+Z_{n-2} + 2Z_{n}).
\end{aligned}
\label{eq:discrete_normalstress}
\end{equation}
While the above expression is written for the special case of a single forcing frequency, it may, in general, have multiple components. In what follows, we shall represent by 
$\mathcal{L}(Z_{n,o})$
a linear function of $Z_{n,o}$, where the 
subscripts represent the appropriate mode $n$ and the corresponding offset $o$. The offset for the above simple case happens to be 1, -1 and 0.
Upon further simplification by substituting the solution \eqref{eq:solution_W} 
into the above equation, we obtain the following:
\begin{gather*}
(T_1 + T_2 + T_3  + T_4 +T_5)_n Z_n =\gamma V_0^2   \mathcal{L}(Z_{n,o}),\\
T_1 = \eta_1 (-q_n^2 + k^2) (Q_{1,n}k+S_{1,n}q_n) + 
\eta_2 (-r_n^2 + k^2) (P_{2,n}k+R_{2,n}r_n), \\
T_2 = (\eta_1S_{1,n}q_n^3 + \eta_2R_{2,n}r_n^3) - k^2(\eta_1S_{1,n}q_n + 
\eta_2R_{2,n}r_n)\\
T_3 = -2\eta_1k^2(Q_{1,n}k + S_{1,n}q_n)\\
T_4 = (\rho_1 - \rho_2)gk^2,\\
T_5 = -\sigma k^4, \\
\gamma = 4\epsilon_1k^3/h_1^2,
\label{eq:general_eigenvalue}
\end{gather*}
which therefore casts the problem in the form of a generalized eigenvalue problem where 
the eigenvalue is the amplitude of the voltage squared.
Mathematically, we have
\begin{eqnarray}
\frac{1}{\gamma}A Z = V_0^2 B Z,
\label{eq:matrix_eigenval}
\end{eqnarray}
where the matrix $A$ (where each row of $A$ is a sum of $T_i$, $i = 1, ..., 5$) is a diagonal matrix and matrix $B$ is a banded matrix 
depending on the input signal function; $Z$ represents the column vector of the 
various coefficients $Z_n$. 
For example, for the case of the single applied 
frequency, we have the following structure 
\begin{equation}
AZ = \left( {\begin{array}{*{20}{c}}
   {{A_{ - 2}}} & {} & {} & {} & {}  \\
   {} & {{A_{ - 1}}} & {} & {} & {}  \\
   {} & {} & {{A_0}} & {} & {}  \\
   {} & {} & {} & {{A_1}} & {}  \\
   {} & {} & {} & {} & {{A_2}}  \\

 \end{array} } \right)\left( {\begin{array}{*{20}{c}}
   {{Z_{ - 2}}}  \\
   {{Z_{ - 1}}}  \\
   {{Z_0}}  \\
   {{Z_1}}  \\
   {{Z_2}}  \\

 \end{array} } \right),BZ = \left( {\begin{array}{*{20}{c}}
   2 & 1 & {} & {} & {}  \\
   1 & 2 & 1 & {} & {}  \\
   {} & 1 & 2 & 1 & {}  \\
   {} & {} & 1 & 2 & 1  \\
   {} & {} & {} & 1 & 2  \\

 \end{array} } \right)\left( {\begin{array}{*{20}{c}}
   {{Z_{ - 2}}}  \\
   {{Z_{ - 1}}}  \\
   {{Z_0}}  \\
   {{Z_1}}  \\
   {{Z_2}}  \\

 \end{array} } \right),
\end{equation}
where the above representation is for $n = 2$ which yields a total of 
$2n+1 = 5$ unknown coefficients. 
It is clear from the above discussion that if the applied voltage has multiple 
modes where the highest mode has a frequency $m\omega$, the matrix $B$ is 
banded matrix with a bandwidth $2m+1$. 
The other thing to note regarding the structure of the matrix $B$ is that for multiple imposed modes, the banded structure will contain all possible combinations of offsets. This implies that for imposed voltage modes $f_1, f_2, ..., f_n$ (corresponding to integers), the offsets will be of the general form $f_i \pm f_j$, $(i,j)\in [1,n]$ with the maximum and minimum width being $\pm 2\max(f_i)$ respectively. 

\subsection{Finite height and its discrete representation}

While the above approximation of large electrode separation leads to comparatively simple expressions for the constants describing the various velocity modes, we would like to pursue the more general case of finite heights. The assumption of large electrode separation implies that to achieve a given threshold electric field, one would have to apply a proportionately large voltage. 
 In order to obtain the appropriate discrete representation of the eigenvalues as done in the previous section, we must begin with the complete form of the solutions for the modes of the vertical velocities, i.e. equation 
\eqref{eq:W_comp_sol}. 
The coefficients of $W_{1,n}$ and $W_{2,n}$ are found out by solving the 
following set of linear equations obtained by substituting equation \eqref{eq:W_comp_sol} into equations \eqref{eq:bcs_1} to \eqref{eq:bcs_2}
\begin{gather}
P_{1,n}\exp{(kh_1)} + Q_{1,n}\exp{(-kh_1)} + R_{1,n}\exp{(q_nh_1)} + 
S_{1,n}\exp{(-q_nh_1)} = 0, \\
kP_{1,n}\exp{(kh_1)} -k Q_{1,n}\exp{(-kh_1)} + q_nR_{1,n}\exp{(q_nh_1)} - 
q_nS_{1,n}\exp{(-q_nh_1)} = 0, \\
P_{2,n}\exp{(-kh_2)} + Q_{2,n}\exp{(kh_2)} + R_{2,n}\exp{(-r_nh_2)} + 
S_{2,n}\exp{(-r_nh_2)} = 0, \\
kP_{2,n}\exp{(-kh_2)} - kQ_{2,n}\exp{(kh_2)} + r_nR_{2,n}\exp{(-r_nh_2)} - 
r_nS_{2,n}\exp{(-r_nh_2)} = 0, \\
P_{1,n} + Q_{1,n} + R_{1,n} + S_{1,n} = P_{2,n} + Q_{2,n} + R_{2,n} + S_{2,n}, 
\\
kP_{1,n} -k Q_{1,n} + q_nR_{1,n} -q_n S_{1,n} =k P_{2,n} -kQ_{2,n} + r_nR_{2,n} 
- r_n S_{2,n},
\end{gather}
\begin{equation}
\begin{split}
\eta_1 \left( 2k^2P_{1,n}+2k^2Q_{1,n} + k^2(S_{1,n}+R_{1,n})+q_n^2R_{1,n} + 
q_n^2S_{1,n}\right) \\= \eta_2 \left( 2k^2P_{2,n}+2k^2Q_{2,n} + 
k^2(S_{2,n}+R_{2,n})+r_n^2R_{2,n} + r_n^2S_{2,n}\right)
\end{split}
\end{equation}
\begin{gather}
-\nu_1\left( -q_{n}^2 + k^2 \right)Z_n = P_{1,n}+Q_{1,n}+R_{1,n}+ S_{1,n}.
\end{gather}
The solution of the above set of equations yields the coefficients ${P,Q,R,S}_{i,n}$ 
($i=1,2$) in terms of $Z_n$.
Upon knowledge of the coefficients from the solution of the linear set of equations, we can substitute the velocities $W_{1,n}$  and $W_{2,n}$ into equation \eqref{eq:discrete_normalstress} to obtain an equivalent representation of equation \eqref{eq:matrix_eigenval}. For completeness, we indicate the terms $T_{1,2,3}$, keeping in mind that the terms $T_4$ and $T_5$ are the same in both the cases of finite and large electrode separation. 
\begin{equation}
\begin{split}
T_1 = (\eta_1(-P_{1,n}+Q_{1,n})+\eta_2(P_{2,n}-Q_{2,n}))k^3 +(-q_n(R_{1,n}-S_{1,n})
\eta_1 + r_n \eta_2(R_{2,n}-S_{2,n}))k^2 + \\
(q_n^2(P_{1,n}-Q_{1,n})\eta_1 - r_n^2
\eta_2 (P_{2,n}-Q_{2,n}))k+q_n^3(R_{1,n}-S_{1,n})
\eta_1 -r_n^3\eta_2 (R_{2,n}-S_{2,n})
 \end{split}
 \end{equation}
 \begin{eqnarray}
T_2 =q_n(k^2-q_n^2)(R_{1,n}-S_{1,n})\eta_1 -r_n(k^2-r_n^2)(R_{2,n}-S_{2,n})\\
T_3 = 2(\eta_1-\eta_2)k^2(k(P_{1,n}-Q_{1,n}) + q_n(R_{1,n}-S_{1,n}))
\end{eqnarray}

\subsection{Limit of ideal fluids}
\label{sec:finite_h}
Later on, it will be convenient to make a comparison of the formulation developed above with the limit of ideal fluids \citep{yih1968stability}.
We begin by dropping the 
viscosities in the governing equation for the $z$-component of the velocity. 
Thus, we must solve the following equations for the vertical velocity modes in an inviscid fluid. 
\begin{eqnarray}
 (s+i(\alpha + n)\omega) \left( \frac{\partial^2}{\partial z^2} - 
k^2\right)W_{1,n} = 0, \\
 (s+i(\alpha + n)\omega) \left( \frac{\partial^2}{\partial z^2} - 
k^2\right)W_{2,n} = 0, 
\end{eqnarray}
which are now subjected to the following boundary conditions: 
\begin{eqnarray}
 W_{1,n}(h_1) = 0, \; W_{2,n}(-h_2) = 0,\\
 W_{1,n}(0) = W_{2,n}(0), \\
 (s+i(\alpha + n)\omega) Z_n = W_{1,n}(0),
\end{eqnarray}
\begin{equation}
 \begin{aligned}
   (\rho_1 - \rho_2)\frac{\partial^2 w_1}{\partial z\partial t} = 
-(\rho_1-\rho_2)gk^2\sigma \zeta+ \sigma k^4\zeta- 
\epsilon_1\frac{4V^2}{h_1}k^3\coth 
kh_1 \zeta.
 \end{aligned}
\end{equation}
Quite obviously, the no-slip condition has been dropped from the boundary conditions.
By assuming that the initial vorticity is zero, we may write the solution for 
the velocity field in the following manner
\begin{eqnarray}
 W_{1,n} = A_1\exp (-kz) + B_1 \exp (kz), \\
 W_{2,n} = A_2\exp (-kz) + B_2 \exp (kz).
\end{eqnarray}
Let us first simplify matters by assuming a large electrode separation, i.e. $h_1, h_2 \rightarrow \infty$. This allows us to write
\begin{equation}
 W_{1,n} = A_1\exp (-kz), \qquad  W_{2,n} = B_2 \exp (kz).
\end{equation}
Utilizing the boundary condition $W_{1,n}(0) = W_{2,n}(0)$, we obtain $A_1 
= B_2 = C_n \text{(say)}$.
Proceeding, we may write the left-hand side of the normal stress balance in the 
form
\begin{eqnarray}
 -(\rho_1 + \rho_2) k(s+i(\alpha + n)\omega)C_n
\end{eqnarray}
which, when employed together with the kinematic constraint at the interface, 
yields the following equation governing the interface deformation
\begin{eqnarray}
 (s+i(\alpha + n)\omega)^2Z_n + \left( \frac{\sigma k^3 - 
(\rho_1-\rho_2)gk}{\rho_1 + \rho_2}-\frac{4\epsilon_1V_0^2k^2}{2(\rho_1 + 
\rho_2)h_1^2}\right) \mathcal{L}Z_{n,o} = 0,
\end{eqnarray}
which is the same equation as that obtined by \cite{yih1968stability} for 
the case of ideal fluids.
We can further simplify the equation obtained above by introducing the following substitutions (please refer to \cite{yih1968stability})
\begin{equation}
p = \frac{\omega_0^2 - \beta}{\omega^2},\qquad 
q = \frac{\beta}{2\omega^2},
\label{eq:yih_22}
\end{equation}
where 
\begin{equation*}
\omega_0^2 = \frac{\sigma k^3 - 
(\rho_1-\rho_2)gk}{\rho_1 + \rho_2},\qquad
\beta = \frac{4\epsilon_1V_0^2k^2}{2(\rho_1 + 
\rho_2)h_1^2}.
\end{equation*}

The corresponding equations for finite heights may be obtained in a similar manner and are not shown here for brevity. The only notable changes are the presence of $\coth kh_1$ in $\beta$ and $\rho_1\coth kh_1 + \rho_2 \coth kh_2$ instead of $\rho_1 + \rho_2$ in the above equations.

\section{Results and discussion}

 We shall first dwell upon the case of a large separation between the electrodes, i.e. $kh_1 \gg 1$. Later on, we shall study finite heights. For understanding the behaviour of such a system, it is sufficient to consider the former assumption  of large separation to be valid. In what follows, we choose aqueous salt solution (e.g. KCl) as the bottom fluid and castor oil as the top fluid. The physical properties are as follows: $\rho_1 = 961$ kg/m$^3$, $\eta_1 = 30\times 10^{-3}$ 
Pa.s,
 $\epsilon_1 = 4\epsilon_0$ (where $\epsilon_0$ represents the dielectric 
permittivity of vacuum),  $\rho_2 = 1000$ kg/m$^3$, $\eta_2 = 8.9\times 
10^{-4}$ Pa.s, $h_1 = 1$ cm, $\sigma$ = $30\times 10^{-3}$ N/m. We note that despite assuming a large height for the hydrodynamics, we must specify a height for the electrostatic problem, i.e. to specify an electric field $V(t)/h_1$. 

In what follows, we shall first look at the influence of viscosity towards altering the marginal stability curves. We shall make a comparison with the case of an inviscid fluid. We shall then look at the influence of the frequency towards altering the marginal stability curves. After that, we study the behavior of the critical voltage (the lowest voltage in the marginal stability curve) and the critical wavelength (the wavelength which corresponds to the critical voltage) as a function of the viscosity. We then study the influence of multiple input frequencies on such a system, with a special focus on the appearance of subharmonic tongues. Subsequently, we illuminate the effects of the confinement due to the electrode spacing on the critical parameters. The effect of the finite electrode spacing on the marginal stability curves and the corresponding dependence of the critical parameters will be depicted. 

\begin{figure}
\centering
\includegraphics[trim={0 0cm 0 0cm}, clip, scale=0.5]{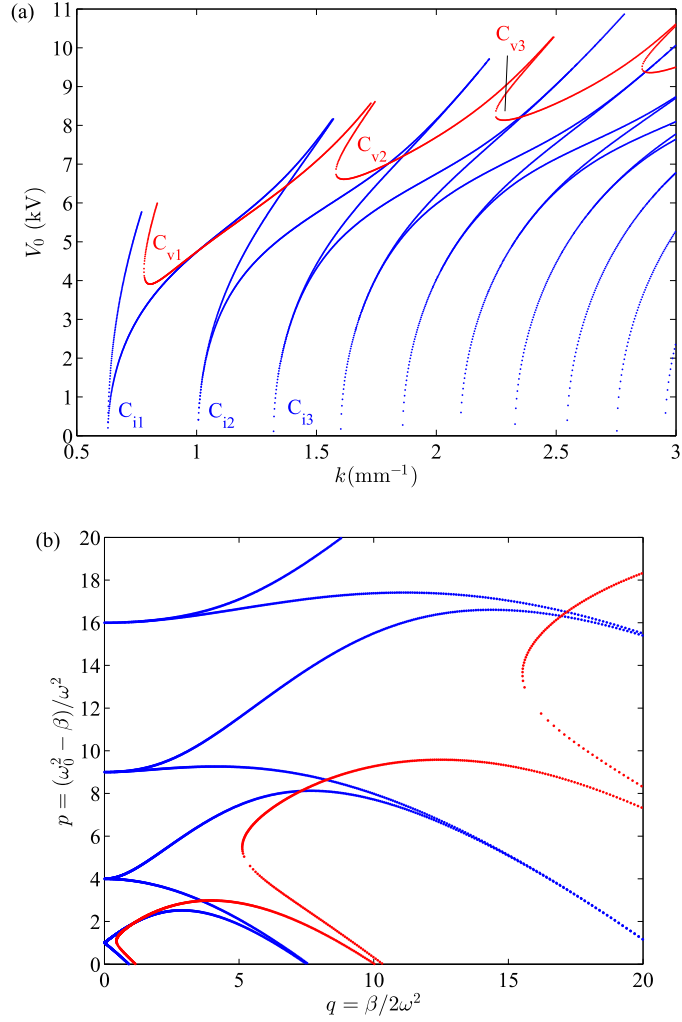}
\caption{(a) Marginal stability curve in the $V_0$-$k$ space for the inviscid 
case (blue curve) and for the case of viscous fluids (red curve). (b) The same 
curves in the $p$-$q$ space. The inviscid case is essentially the marginal stability curve 
predicted by the Mathieu equation (see \cite{yih1968stability}). The fluid 
properties are as follows: $\rho_1 = 961$ kg/m$^3$, $\eta_1 = 30\times 10^{-3}$ 
Pa.s, $\epsilon_1 = 4\epsilon_0$ ($\epsilon_0$ represents the dielectric 
permittivity of vacuum),  $\rho_2 = 1000$ kg/m$^3$, $\eta_2 = 8.9\times 
10^{-4}$ Pa.s, $h_1 = 1$ cm, $\sigma$ = $30\times 10^{-3}$ N/m. The frequency of the applied voltage is $\omega = 20\pi$ s$^{-1}$.}
\label{Fig:marginal_standard}
\end{figure}

The solution to the eigenvalue problem is obtained by setting $s = 0$ to obtain the marginal stability curve. The corresponding generalized eigenvalue problem (equation \eqref{eq:matrix_eigenval}) is solved by means of the \textit{eig} function of {\sc{matlab}}. In short, the wavenumbers are varied, and the solution of the eigenvalue problem yields the voltage corresponding to marginal stability. We utilized $n = 30$ in our work but we note that a lower $n$, say $n = 8$, also suffices for the problem. In order to plot the marginal stability curves, the corresponding first few nonzero magnitudes of the potentials (eigenvalues of the eigenvalue problem) are plotted; this fact corresponds to the various voltages depicting the zones of stability and instability for a given wavenumber, as shall be seen below. 

In figure \ref{Fig:marginal_standard}, we depict (a) the marginal stability curve in the $V_0$ vs $k$ space and (b) in the $p$ vs $q$ space. We shall call the former the physical space and the latter the transformed space. The blue curve depicts the variation for an inviscid fluid system, while the red curve depicts the curve for a system of viscous fluids. The properties used to obtain the marginal stability curves are mentioned in the figure caption. 
In the above figures we have utilized the large separation approximation. 
The critical points for the viscous fluids are denoted by $C_{v,j}$ (subscript $v$ represents viscous), while those for the inviscid fluids  are denoted by $C_{i,j}$ (subscript $i$ represents inviscid). The indices $j$ represent the various local minima of the voltages corresponding to each of the stability tongues (curves resemble tongues).
From subplot (a) we observe the following influences of viscosity: The critical voltage significantly increases - this is seen from the shift of the critical points to higher values of voltage magnitude (see points $C_{v,1}$ and $C_{i,1}$). The critical wavenumber shifts to larger values (the critical wavelength decreases) - this is seen from the shift of the critical points to higher values of the wavenumber. 
Each tongue of marginal stability spans across a larger wavenumber range, as observed from the increasing spacing between the critical voltages (compare the range $C_{v,1}-C_{v,2}$ for the viscous fluids and $C_{i,1}-C_{i,2}$ for the inviscid fluids). 
This provides the counterargument to the inviscid theory which predicts that even for a very small applied voltage (nearly zero as a matter of fact, as noted by \cite{yih1968stability}) there is a critical wavenumber which causes the system to become unstable. 
Because of the fact that the system is in fact excited by  $V^2$ , i.e. integral multiples  $2\omega$, instead of $\omega$, there are no subharmonics for this system, i.e. all the modes are harmonic in this particular case. 

In the $p$-$q$ space (depicted in subplot (b)) we see that marginal stability curves observed resemble a modified version of that for the inviscid case, which corresponds to the Mathieu equation. To reiterate, $p$ represents the magnitude of the restoring forces while $q$ represents the magnitude of the forcing terms. 
For the inviscid case, we see that curves tend to $n^2$ ($n$ is an integer) as $q$ tends to zero. These modes are the simple harmonic modes; the concave regions to the right of this are unstable. In fact, the various points $n^2$ in the transformed space for $q=0$ correspond to the points $C_{i,n}$, $n=1,2,...$ in the real space. 
The influence of viscosity follows from the observations in subplot (a) which is to increase the critical value of $\beta$. 
This underlines the fact that a significant voltage is required to induce instability, which is in contrast to the inviscid case. 
An important observation is that the first harmonic, corresponding to the first tongue, is weakly affected by the effects of viscosity in the region of small $q$. For larger $q$ however, the effects are more pronounced. A large $q$ corresponds a larger applied voltage and wavenumber. We also keep in mind that the increment of $k$ also leads to an increase in the parameter $p$.

\begin{figure}
\centering
\includegraphics[scale=0.65]{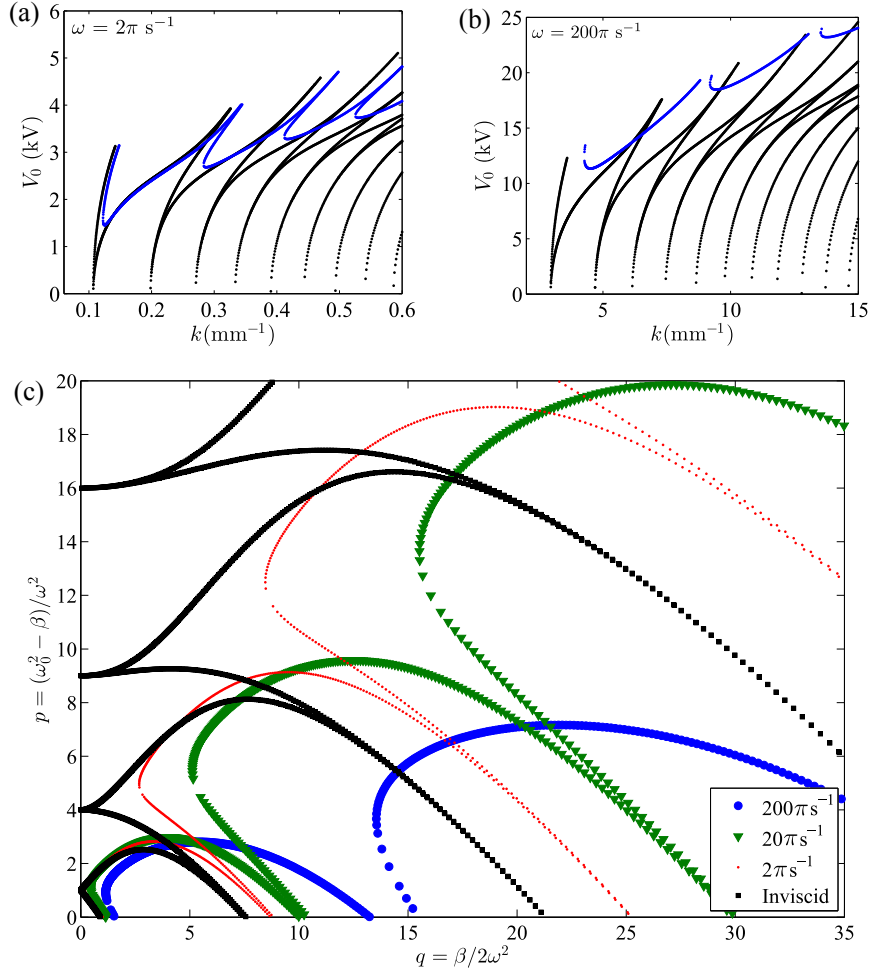}
\caption{Marginal stability curve in the $V_0$-$k$ space for (a) $\omega = 2\pi$ s$^{-1}$ and (b) $\omega = 200\pi$ s$^{-1}$. 
The blue curves represent the case of viscous fluids, while the black curves depict the inviscid case. 
(c) Marginal stability curve in the $p-q$  space. 
The blue, green and red curves represent frequencies of $200\pi$ s$^{-1}$, $20\pi$ s$^{-1}$, and $2\pi$ s$^{-1}$, respectively. The black curve represents the case of an inviscid system. All other physical parameters are the same as reported in figure \ref{Fig:marginal_standard}}
\label{Fig:marginal_stab_diff_freq}
\end{figure}

In figure \ref{Fig:marginal_stab_diff_freq}, we present the effects of various single-frequency applied voltages on the marginal stability curves. Just like in the previous figure, we compare the marginal stability curves in both the physical space and the transformed space. The influence of frequency is immediately seen in the physical space. 
In the subplots, we present the marginal stability curve in the real space for a frequency of (a) $\omega = 2\pi$ s$^{-1}$ and (b) $\omega = 200\pi$ s$^{-1}$. The intermediate case was depicted in figure \ref{Fig:marginal_standard}. We observe that for low frequencies of the applied voltage, the marginal stability curves (and the critical voltage and wavenumber) follow the curve predicted by the inviscid case. The discrepancy in the marginal stability curve is more evident for a larger applied frequency (as seen in subplot (b)). 
It is seen that for a lower applied frequency, the critical wavenumber is low (which corresponds to a large disturbance wavelength). For a higher frequency, the disturbance wavenumbers grow. 
Moreover, an increase in the frequency also leads to a significant increase in the critical voltage. 
A common observation is that the tongues quickly cascade into tongues of increasingly smaller width as we move towards larger wavenumbers. 
The observations on the critical wavenumbers and voltages are better represented in the transformed space where an easy comparison with the marginal stability regime of a Mathieu equation can be done. Interestingly, for the inviscid case, the representation in the $p-q$ space is independent of the applied frequency. 
It can be observed that there is a significant shift in $\beta$  which corresponds to a larger critical voltage as the frequency of the applied voltage increases. 
As was noted in context with subplots (a) and (b), a lower applied frequency resembles the situation for inviscid fluids. 
The observations may be justified by noting that in the equation of the vertical velocity components (equation \eqref{eq:govern_velo_modes}), a low frequency leads to the elimination of the first term. Mathematically, this occurs when $i(\alpha + n)\omega W_{1,n} \ll \nu_1 (d^2/dz^2-k^2)^2 W_{1,n}$. We note that the dropping of the first term is equivalent to neglecting the temporal term in the Navier Stokes equation. Therefore, for the low frequency case, the equations are similar to those where the viscosity is absent. 

\begin{figure}
\centering
\includegraphics[scale=0.6]{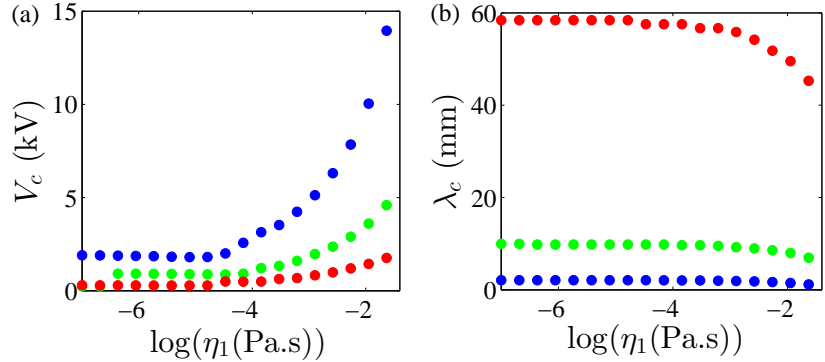}
\caption{Variation of (a) the critical voltage and (b) the critical wavelength with the viscosity of the two fluids. The viscosities of the two fluids are considered to be equal. All  other properties are the same as those considered in figure \ref{Fig:marginal_standard}. The blue, green and red curves represent $\omega$ = 100 Hz, 10 Hz and 1 Hz respectively. }
\label{Fig:critical_eta}
\end{figure}

Figure \ref{Fig:critical_eta} depicts the variation of the critical voltage ($V_c$) and the critical wavelength ($\lambda_c = 2\pi/k_c$) as a function of the viscosity of the top fluid, $\eta_1$. In that case, we have chosen the viscosities of the two fluids to be equal, as this simplification does not obfuscate any of the underlying physics. 
It is observed that for large values of the viscosity, we have a larger critical voltage. 
This may be attributed to the fact that a larger viscosity acts as a larger damping factor to the forcing electrical stress. 
For smaller viscosity, there is a saturation in the critical voltage, with the critical voltage being higher for a larger applied frequency. 
It may also be observed in subplot (b) that a larger applied frequency has a smaller critical wavelength. This is also seen in figure \ref{Fig:marginal_stab_diff_freq}. As the viscosity is increased beyond a certain point, the critical wavelength decreases sharply. This implies that for high frequency modes, an approximation of long waves would become inappropriate for analyzing the behaviour of the system. 
The critical voltage is higher while the critical wavelength is lower for a higher frequency.

\begin{figure}
\centering
\includegraphics[trim={0 0cm 0 0}, clip, scale=0.65]{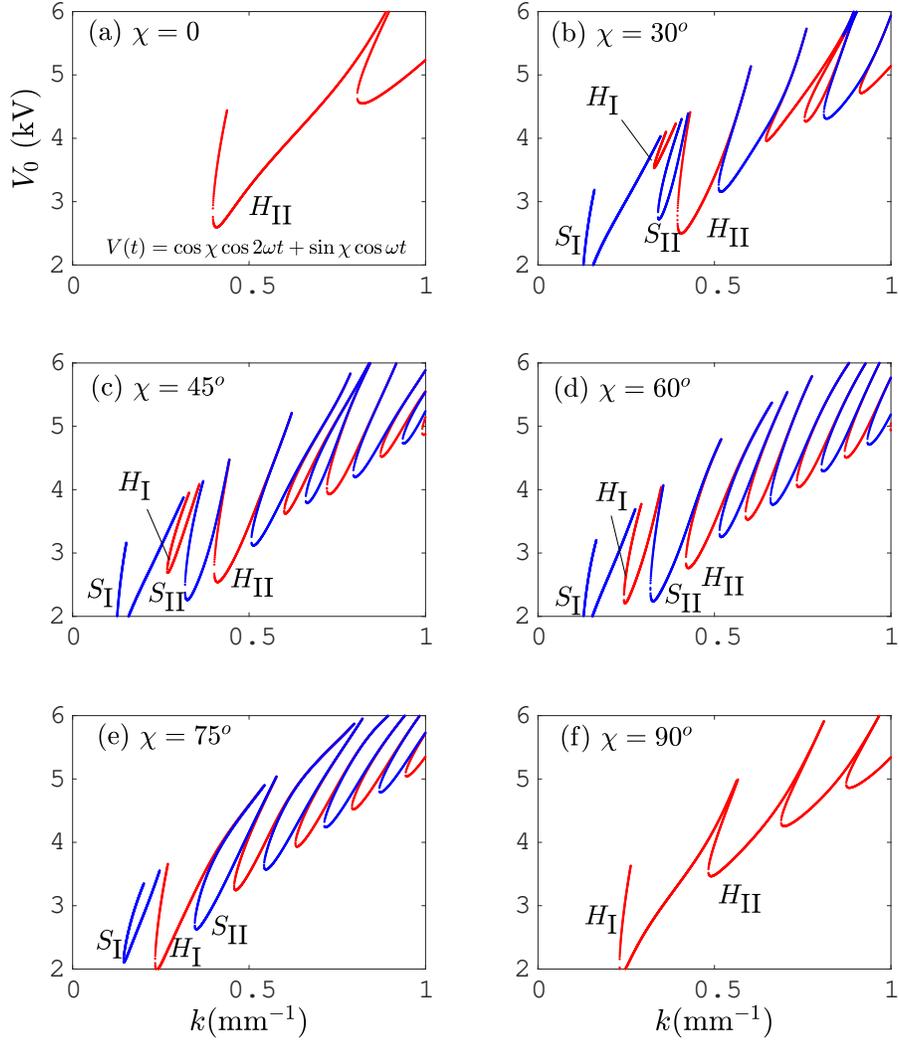}
\caption{Marginal stability curves in the $V_0-k$ space for different mixing factors $\chi$ and $f_1 = 2$, $f_2 = 1$. The subplots (a) and (f) correspond to $\chi = 0^o$ and $\chi = 90^o$, respectively, representing a frequency of $2\omega$ and $\omega$, respectively. An intermediate $\chi$ represents a mixture of these modes. All the other properties are the same as in figure \ref{Fig:marginal_standard} for $\omega = 4\pi$ s$^{-1}$. The tongues denoted by $S_\textrm{I, II}$ and $H_\textrm{I,II}$ represent the subharmonics and harmonics, respectively (subscript $I$ for the first and so on). The general form of the potential considered is displayed in subplot (a). }
\label{Fig:multimode_2_1}
\end{figure}

\begin{figure}
\includegraphics[scale=0.6]{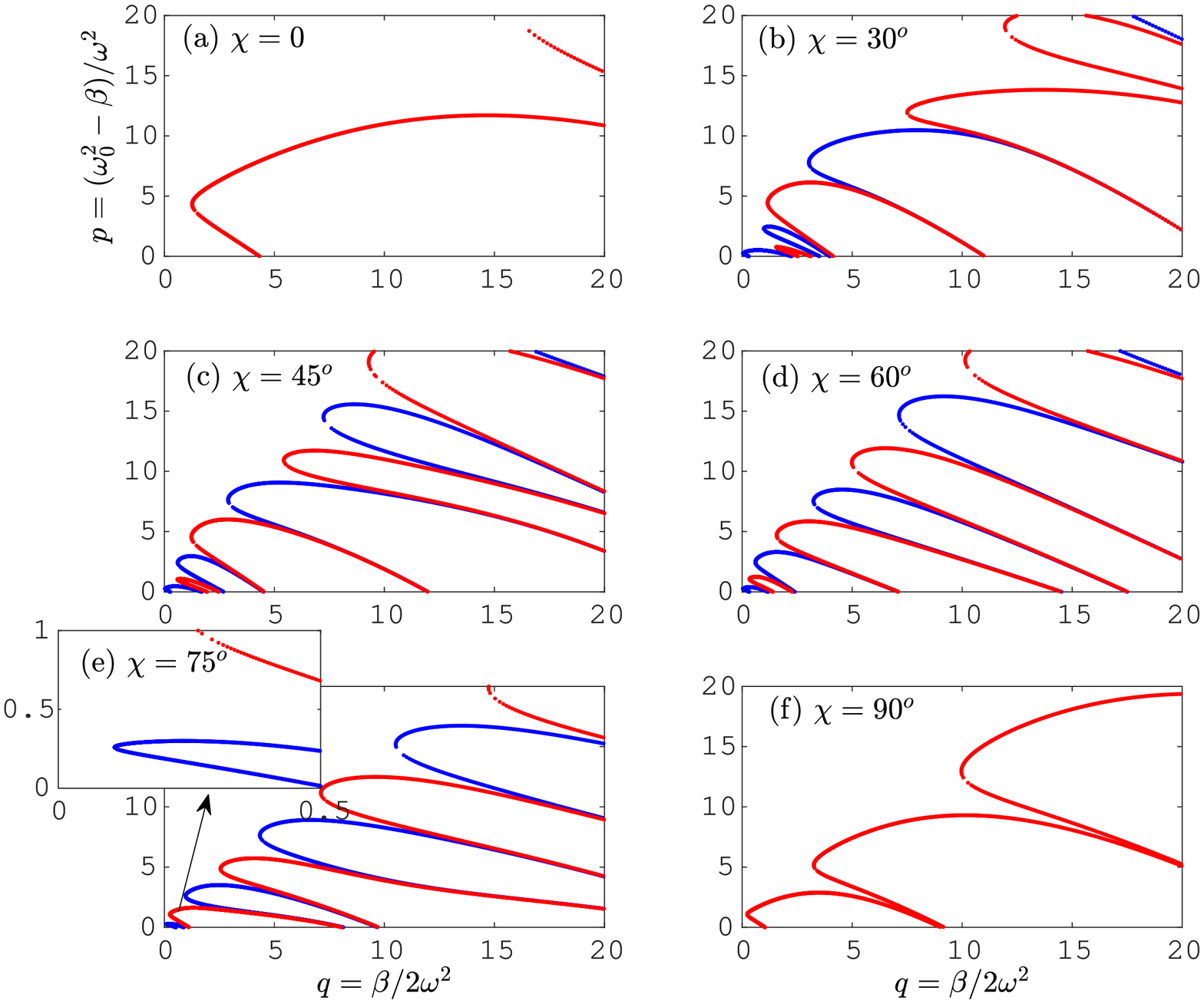}
\caption{Marginal stability curves in the $p-q$ space for different mixing factors $\chi$ and $f_1 = 2$, $f_2 = 1$. The red (blue) tongues are the harmonic (subharmonic) tongues. The corresponding diagram in real space is depicted in figure \ref{Fig:marginal_standard}.}
\label{Fig:p-q-multiple-freq}
\end{figure}

Having discussed the influence of a single frequency on the marginal stability curve, we turn our attention to the general case of mixed modes. Towards this, we consider a potential of the form 
\begin{eqnarray}
V(t) = \cos \chi \cos f_1\omega t + \sin \chi \cos f_2 \omega t
\end{eqnarray}
where $\chi$ represents the mixing factor as defined by \cite{besson1996two}. The two parameters appearing above, $f_1$ and $f_2$, are integers and represent the simple scenario of a binary mix of two frequencies. 
We must now determine how multiple modes are captured in the generalized eigenvalue problem framwork discussed in equation \eqref{eq:general_eigenvalue}.
To see this, we first square the above expression for the time-periodic electric potential to obtain
\begin{equation}
\begin{split}
V(t)^2 = \frac{\cos^2\chi}{4}
\left(
\exp(2f_1i\omega t) + \exp(-2f_1i\omega t)
\right)
+
\frac{\sin^2\chi}{4}
\left(
\exp(2f_2i\omega t) + \exp(-2f_2i\omega t)
\right)
+\\
\frac{\sin 2\chi}{8}\left(
\exp(2(f_1+f_2)\omega t) + \exp(-2(f_1+f_2)\omega t) + \exp(2(f_1-f_2)\omega t) + \exp(2(f_2-f_1)\omega t) 
\right)+ \frac{1}{4}
\end{split}
\end{equation}
which we can then multiply by the mode $Z_n$. By performing the shift in indices we obtain nonzero entries for the following columns of the $n$th row of the matrix $B$: $n\pm 2f_1$, $n\pm 2f_2$, $n\pm 2(f_1+f_2)$, $n\pm 2(f_1-f_2)$, and $n$. 
All indices except $n$ represent nondiagonal entries to the matrix $B$. 
As $\chi$ is varied from $0^o$ to $90^o$, we perform the transition from a frequency $f_1$ to a frequency $f_2$. Clearly, it is the intermediate situation that is of interest to us.

The case $\chi = 0$, corresponding to a single frequency $f_1$, is depicted in figure \ref{Fig:multimode_2_1} (a). 
It is observed that like in the previous case, there is only the contribution from the single mode, i.e. all the modes are harmonic, as attributed to the $V(t)^2$ nature of the driving force. The first point, which is the critical point for $\chi = 0^0$, is represented as $H_\textrm{II}$ because it will be seen that this point eventually becomes the second harmonic tongue for $\chi = 90^o$. 
An increase in the mixing factor leads to the appearance of subharmonic modes, as observed for subplots (b) through (e) (with increments of $15^o$). 
The subharmonic tongues are represented by $S_\textrm{I}$.
The origin of such subharmonic modes is due to the presence of odd terms in the mixed modes of $V(t)^2$ as obtained above. 
Consequently, the blue modes appearing for nonzero $\chi$ are subharmonic, i.e. they have a frequency which is a multiple of $\omega/2$ \citep{kumar1994parametric}.
 This implies that now the most unstable wavenumber does not correspond to a response with frequency $\omega$, but rather half of that. 
As the mixing factor is further increased, we observe that the small harmonic tongue $H_\textrm{I}$ that appears in a region below the previously most unstable wavenumber grows continuously, while the harmonic tongue $H_\textrm{II}$ shrinks in size. Eventually, the first subharmonic tongue, $S_\textrm{I}$, which grows until $\chi = 45^o$ starts to shrink. At $\chi = 90^0$, $S_\textrm{I}$ has vanished completely, and $H_\textrm{I}$ becomes the most unstable mode. 

In figure \ref{Fig:p-q-multiple-freq} we represent the marginal stability curves in the transformed space for the various mixing factors. The key point to be observed here is  once again the fact that for the limiting values of $\chi = 0^o$ and $90^o$, the marginal stability curves exhibit only harmonic frequencies which essentially imply that the disturbance will be oscillating with a integral frequency of $2\omega$ for $\chi = 0^o$ and a frequency of $\omega$ for $\chi = 90^o$. 
For the intermediate values of the mixing factor, one observes, as observed in the previous figure, that there are small zones of subharmonic tongues. These frequencies are integral multiples of $\omega /2$. In the unmixed cases of $\chi = 0^o$ and $90^o$, we see that the lowermost curve approach the \textit{inviscid} limit of $n^2$ (where $n$ is an integer). For $\chi = 0$, the curve is closer to 4 (for an inviscid case, i.e. the Mathieu equation, the critical value goes exactly to 4 for $q = 0$). 
For $\chi = 0$ (which is the case with an applied voltage of frequency of $2\omega$) gravest mode therefore corresponds to a frequency of $2\omega$, the next mode is $4\omega$ and so on. 
For $\chi = 90^o$ (which corresponds to an applied voltage of $\omega$), the first mode is closer to $1$ then $4$ and so on which indicates that first mode is $\omega$, the second is $2\omega$ and so on. The subharmonic modes (which are evaluated for $\alpha = 1/2$) oscillate with modes of frequency $\omega/2$, $3\omega/2$ and so on \citep{benjamin1954stability}.

\begin{figure}
\centering
\includegraphics[scale=0.6]{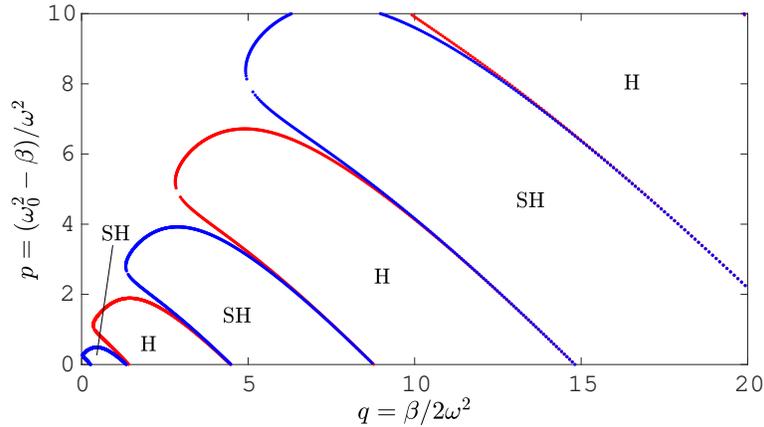}
\caption{Marginal stability curve in the $p-q$ space for an ac field biased by a dc field: $V(t) = \frac{1}{\sqrt{2}}(1 + \cos \omega t)$, $\omega = 4\pi$ s$^{-1}$. All other properties are the same as in figure \ref{Fig:marginal_standard}. }
\label{Fig:DC_bias}
\end{figure}

\begin{figure}
\centering
\includegraphics[scale=0.6]{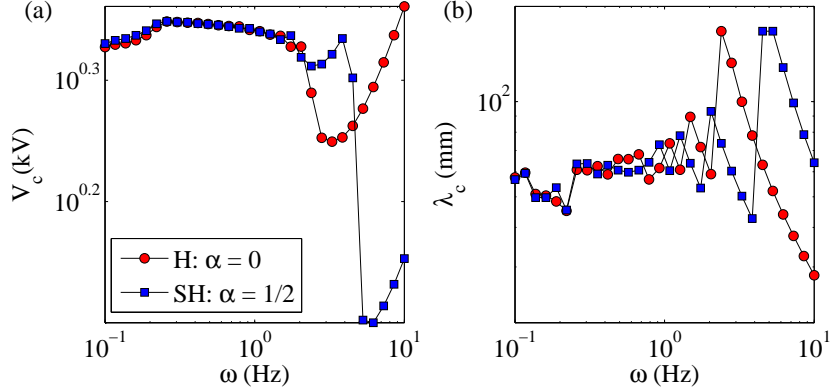}
\caption{Critical (a) voltage (a) and wavelength (b) as a function of the ac frequency of an applied voltage of the form $V(t) = \frac{1}{\sqrt{2}}(1 + \cos \omega t)$. All other parameters are the same as in figure \ref{Fig:marginal_standard}. The circular markers (red) and the square markers (blue) represent the harmonic and subharmonic response, respectively. $\alpha = 0$ and $1/2$ represent the critical quantities calculated for the harmonic and subharmonic tongues respectively.}
\label{Fig:critical_biased}
\end{figure}

As a natural extension of the ideas mentioned above, we can set the frequency of one of the modes to zero. This is equivalent to having a dc bias on an ac field. The resulting stability curves are is depicted in figure \ref{Fig:DC_bias}. The mixing factor is chosen as $\chi = 45^o$, and the other frequency is chosen to be $\omega$, i.e. the form of the applied voltage is $V(t) = \frac{1}{\sqrt{2}}(1 + \cos \omega t)$.  
The squared form of $V(t)$ yields frequencies of $2\omega$, $\omega$ and $0$. 
In this manner, one does not only obtain harmonics, (see the subfigure (f) of figure \ref{Fig:multimode_2_1} which represents the case of a single frequency $\omega$) but also  subharmonics, in an alternating manner.
The alternating subharmonic and harmonic tongues are reminiscent of the Mathieu equation, except for the effect of viscosity which causes the marginal stability curves to shift to higher values of $\beta$ (which correspond, as noted earlier, to nonzero voltages). 
In figure \ref{Fig:critical_biased} we depict the corresponding critical quantities as a function of the applied ac frequency. It is observed that the effect of frequency on this system is much pronounced for higher frequencies. The shifting of the marginal curves happens in such a manner that first the harmonics are the most unstable modes (i.e. the critical voltage for the harmonic case is lower than that for the subharmonic one), while an increase in frequency causes a cross-over after which the subharmonics become the most unstable modes. The effect of this on the most unstable wavenumber is also depicted in subplot (b) which shows multiple cross-overs at larger frequencies.

\begin{figure}
\centering
 \includegraphics[trim={1cm 0 0 0},clip,scale=0.55]{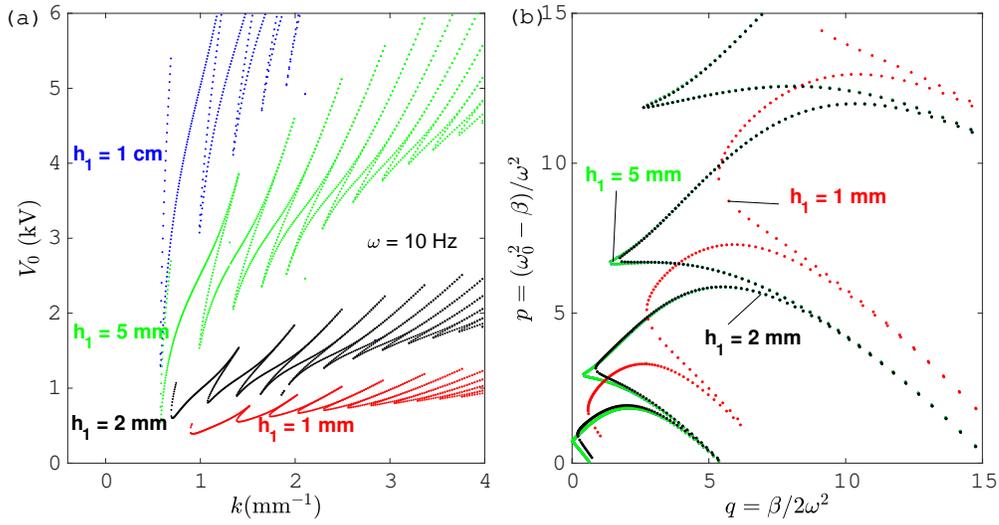}
 \caption{Marginal stability curves in the (a) $V_0 -k$ space and (b) $p-q$ space for systems with finite electrode spacing. The red curve depicts the case with $h_1 = h_2 = 1$ mm, the black curve $h_1 = h_2 = 2$ mm, the green curve $h_1 = h_2 = 5$ mm and the blue curve $h_1 = h_2 = 1$ cm. The frequency is chosen to be $20\pi$ s$^{-1}$. }
 \label{Fig:finite_h}
\end{figure}

 Figure \ref{Fig:finite_h} depicts the marginal stability curves for the case where the electrodes are separate by a finite distance. 
Subplot (a) depicts the stability curve in the $V_0-k$ space while (b) depicts the same in the $p-q$ space. 
The curves have been evaluated for $h_1 = $1 cm, 5 mm, 2 mm, and 1 mm at a frequency of 20$\pi$ s$^{-1}$. 
 A reduction in electrode spacing results in a lowering of the threshold voltage for a given wavenumber, which may be explained by noting that a decrease in the distance between the interface and electrode leads to a larger electric field  in the first fluid for a given applied voltage. 
 In particular, we see that the reduction in the critical voltage leads to an upward shift of the stability tongue in the $p-q$ plane. 
$q$ increases primarily because of the increase in the wavenumber. The decrease in the critical voltage, which corresponds to the lowering of the electrode spacing, plays a minor role in the magnitude of the parameter $q$.
It must, however, be kept in mind that the parameter $\omega_0$ decreases significantly as the height decreases (owing to the contribution from $\coth kh_2$ terms in the denominator; please see equation 
\eqref{eq:yih_22}).  
 The other implicit effect of a finite gap is due to the viscous stresses mediated through the no-slip condition. 
 This means that if the characteristic viscous penetration length is smaller than the distance between the deforming interface and electrode wall, the effect of the wall is felt very weakly. 
 In such a case, the characteristic length due to the viscous forces may be written as $l_\text{vis} = \sqrt{\nu_1/\omega}$, where $\nu_1$ is the kinematic viscosity. 
 Considering a system composed of an aqueous KCl solution and castor oil, we see that this length turns out to be $\approx$ 2.2 mm for an applied frequency $\omega = 1$ Hz. 
 As the frequency is increased, the characteristic viscous length decreases. An increase in the frequency therefore implies that the finite length will not have an immediate impact on the system unless the spacing is reduced down to the order of that viscous lengthscale. 
 In particular, for a frequency of  100 Hz, we find that $l_\text{vis} = 0.23$ mm. While there is the explicit alteration of the critical voltage due to reduction in height, the conversion into the $p-q$ space reveals that the larger heights are not much responsive to the applied frequency, in line with the above discussion. 
 This fact is observed in subplot (b) where we see the alteration of the marginal stability curve for various distances between the electrode and the fluid interface of the two fluids (red, black, green, and  blue curves represent 1, 2, 5, 10 mm). 
 It is seen that for heights of 1 cm and 5 mm, there is no appreciable alteration in the marginal stability curve. However, when we approach smaller lengths, we observe that there is an increase in $\beta$ and $p$. $\beta$ depends on $V_0^2k^2/h_1^2$, which implies that given an equal field strength (which is approximately constant as seen from subplot (a)), an increase in the critical wavenumber $k$ (as seen from subplot (a)) leads to an increase in $\beta$ and thus $q$. 

\begin{figure}
\includegraphics[trim={0.3cm 0 0 0},clip,scale=0.6]{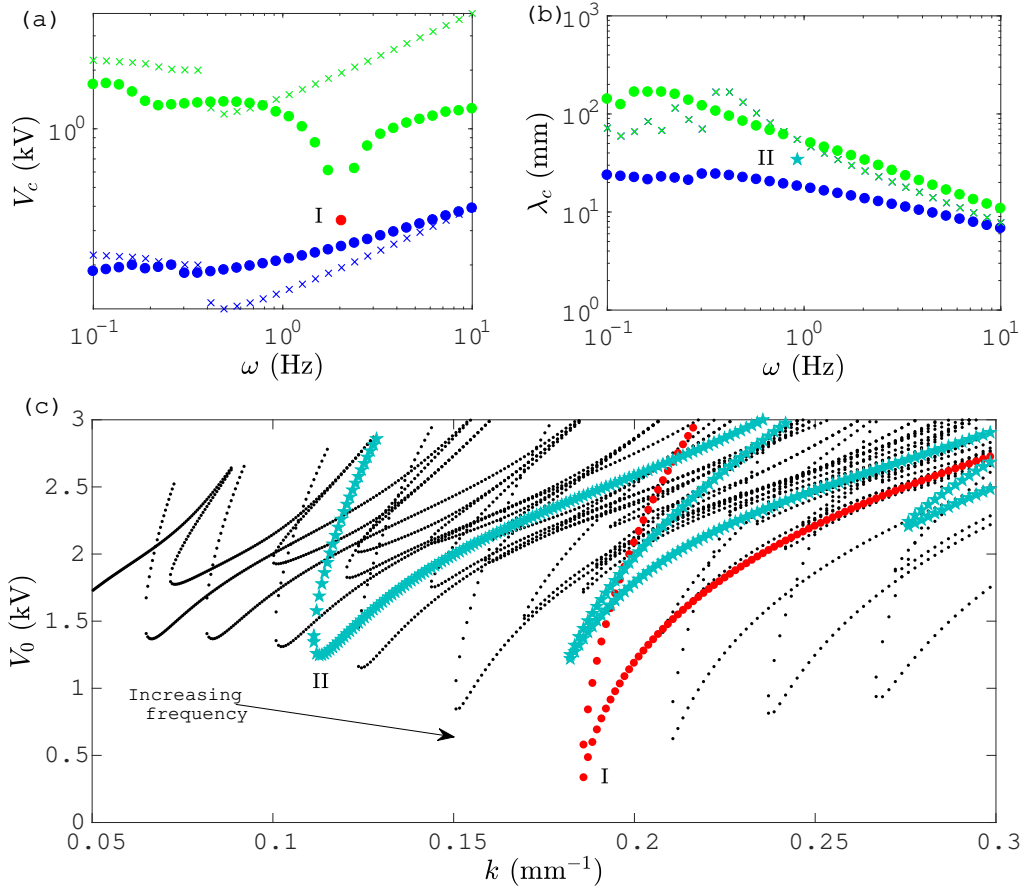}
\caption{(a) Critical voltage and (b) critical wavelenght as a function of the applied frequency. All the other properties are the same as those in figure \ref{Fig:marginal_standard}. The blue and green symbols represent the case of $h_1$ = 1 mm and 1 cm, respectively. The filled markers represent the calculations done with finite electrode spacing, while the crosses represent those done with the infinite height approximation. (c) Marginal stability curves in the $V_0 - k$ space for various frequencies and $h_1 = 1$ cm. The leftmost curve represents the lowest frequency. A frequency increase shifts the curves to the right.s going right to left represent those due to decreasing frequencies. The red curve represents the marginal stability curve for which the critical voltage is minimal for $h_1$ = 1 cm (the red filled marker in subplot(a)). The marginal curve depicted by the star marker is that for which there are two wavenumbers corresponding to the same critical voltage. This indicates the possibility of bistability. }
\label{Fig:critical_params_vs_freq}
\end{figure}

In figure \ref{Fig:critical_params_vs_freq} we depict the variation of the critical voltage and the critical wavelength with the applied frequency.
The results obtained by the infinite-height approximation are depicted by the cross markers, while the solution obtained by accounting for the finite height are depicted by filled markers. 
The blue and green markers represent an electrode spacing of $h_1$ = 1 mm and 1 cm, respectively.  
It is observed that for $h_1 = 1$ mm, there is an insignificant difference between the two approaches towards determining the critical wavelength, especially at high frequencies. 
However, for $h_1 = 1$ cm, the two approaches do not agree with each other. 
Quite remarkably, for larger $h_1$, we observe that there is an optimal frequency for which the critical voltage is the lowest. 
The larger difference at higher frequency may be attributed to the fact that a larger frequency, the temporal term is higher, as was observed in figure \ref{Fig:critical_eta}.  The complicated interplay between the electrode separation and the applied frequency in the presence of finite viscosity leads to the observed variations in the critical voltage and wavelengths. 

If we focus on the local minima of the critical voltage, which is depicted by the red marker (point I) in subplot (a) and the red marginal curve in subplot (c). Subplot (c)  depicts the marginal stability curves for various frequencies. 
The marginal curves for lower frequencies shift to lower wavenumbers (higher wavelengths (as seen in subplot (b)). 
In the event of such a local minimum (denoted by point II, the star marker), there is a frequency at which there are two wavenumbers which yield the same critical voltage, as observed from the marginal curve marked with the star symbols. 
This suggests the possibility of a bistability, a behavior that was observed by  \cite{kumar1996linear}. We shall dwell upon this in a future work.

\section{Conclusions}
To conclude, we have performed a Floquet analysis to study a system of two immiscible viscous fluids with a horizontal interface exposed to an arbitrary time periodic ac electric field. The field is created by parallel electrodes of a finite spacing. 
We derived the discrete representation for the various Fourier modes and obtained a generalized eigenvalue problem arising from Floquet theory. 
For single applied frequencies, we showed that all excitations of the system have harmonic frequencies. The reason is attributed to the normal electric stress at the interface that acts as the excitation source. Because the stress is proportional to the electric field squared, no subharmonic modes are found.
The effect of viscosity is shown to yield a stability threshold with nonzero critical voltages for finite wavenumbers, which is in contrast to the limiting scenario of inviscid fluids (see \citep{yih1968stability}). 
When studying the variation of the critical parameters with viscosity, a plateau of the critical voltage and wavelength for small viscosities is found.  
Furthermore, an input ac voltage with multiple frequencies was studied based on the corresponding generalized eigenvalue problem. 
Interestingly, we find that the cross-coupling of the multiple excitation modes leads to the appearance of a subharmonic response, whereas a single-frequency excitation only yields harmonic modes. Details of the mixing of the excitation modes determine whether the most unstable mode of the system is of  harmonic or subharmonic nature. 
The special case of an ac frequency superposed by a dc bias voltage was also analyzed. 
The effect of the channel height on the critical voltage and wavenumbers is also depicted. 
The effect of finite electrode spacing was shown to be more prominent for lower applied frequencies, which correspond to a thicker viscous boundary layer.  
However, at the same time, we have shown that the a lower applied frequency leads to a smaller temporal variation in the momentum equation, thus reducing the viscous effects. 
The findings reported in this paper shed light on the process of destabilization of a liquid-liquid interface under realistic conditions, especially for nonzero viscosity, a finite electrode spacing, and an arbitrary applied voltage. We therefore hope that this work may prove useful in a number of practical applications such as the electric-field assisted structure formation at fluid interfaces.

\section{Acknowledgments}
AB gratefully acknowledges the Alexander von Humboldt foundation for postdoctoral funding.

\end{document}